\documentclass[twocolumn,,superscriptaddress,letter]{revtex4}%
\usepackage{amssymb}
\usepackage{color}
\usepackage{graphicx}
\usepackage{dcolumn}
\usepackage{bm}
\usepackage[header,title,page,titletoc]{appendix}
\usepackage{amsmath}
\usepackage{amsfonts}%
\providecommand{\U}[1]{\protect \rule{.1in}{.1in}}

\begin{document}
\title{Quantum Phase Liquids - Fermionic Superfluid without Phase Coherence}
\author{Ya-Jie Wu}
\affiliation{Department of Physics, Beijing Normal University, Beijing 100875, China}
\author{Jiang Zhou}
\affiliation{Department of Physics, Beijing Normal University, Beijing 100875, China}
\author{Su-Peng Kou}
\thanks{Corresponding author}
\email{spkou@bnu.edu.cn}
\affiliation{Department of Physics, Beijing Normal University, Beijing 100875, China}

\begin{abstract}
We investigate the two dimensional generalized attractive Hubbard model in a
bipartite lattice, and find a "quantum phase liquid" phase, in which the
fermions are paired but don't have phase coherence at zero temperature, in
analogy to quantum spin liquid phase. Then, two types of topological quantum
phase liquids with a small external magnetic field--$Z_{2}$ quantum phase
liquids and chiral quantum phase liquids--are discussed.

\end{abstract}
\maketitle

\section{Introduction}

In condensed matter physics, Landau's symmetry breaking paradigm had been
considered as the foundation to learn all types of orders. Different (quantum)
states are characterized by different symmetries and the associated local
order parameters. However, in last 20 years, it became more and more clear
that Landau's paradigm cannot describe all quantum states of the matter. The
first example beyond Landau's paradigm is the fractional quantum Hall (FQH)
effect\cite{TSG8259}. The FQH states possess exotic topological orders and
cannot be described by symmetry breaking theory. The subtle structures that
distinguish different FQH states are called\emph{ topological order}%
\cite{wen,Wenrig,Wtoprev}. Another example is the quantum spin liquid, which
is an intriguing possibility for a strongly interacting magnetic system where
the magnetically ordered ground state is avoided owing to strong quantum
fluctuations. The subtle structures that distinguish different quantum spin
liquid states are called\emph{ quantum order}\cite{wen,wen2,Wenpub}. In the
projective space construction of spin liquid with the same (global) symmetry,
there exist many kinds of quantum spin liquids with different low energy gauge
structures ($\mathrm{SU(2)}$, $\mathrm{U(1)}$, $Z_{2}$...). To distinguish
different quantum spin liquids, the projective space symmetry group (PSG) is
proposed\cite{wen,wen2,Wenpub}. People have tried to find such exotic quantum
states in frustrated spin systems for more than twenty years\cite{fazekas}.
For example, the two-dimensional (2D) \textrm{J}$_{1}$\textrm{-J}$_{2}$ model
is a typical frustrated Heisenberg model on a square lattice. Different
methods show a similar result - a quantum disordered ground state appearing
around $J_{2}/J_{1}\sim0.5$. However, the nature of the quantum disordered
ground state is still much
debated\cite{russi,Dagotto,Sano,Schultz,Sorella,cap}.

In this paper, we find that the fermionic superfluid (SF) may lose long range
phase coherence due to quite strong quantum fluctuations. The strongly
fluctuating fermionic SF possesses exotic quantum orders and also cannot be
described by Landau's symmetry breaking theory. We call it \emph{quantum phase
liquid} (QPL). In the QPL, the fermions are paired and single quasi-particle's
excitation has a finite energy gap. However, the strong quantum fluctuations
destroy the long range phase coherence and the fermionic SF order parameter is
still zero. Namely, QPL state is a fermionic SF without phase coherence, and
the excitations may be doublons or holons that correspond to the quantum
states with the particle number $2$ or $0$ on each site, respectively. The
single quasi-particle excitations corresponding to the quantum states with the
particle number $1$ on each site are forbidden. It is found that physical
properties of topological quantum phase liquids are quite different from those
of topological quantum spin liquids--the external magnetic field induces a gas
of topological excitations which will condense at zero temperature in a
topological QPL.

The paper is organized as follows: In Sec. II we introduce the two dimensional
generalized attractive Hubbard model in a bipartite lattice, and discuss the
\textrm{SU(2)} pseudo-spin symmetry of the model. In Sec. III we introduce
projective space construction of QPLs in analogy to quantum spin liquids. In
Sec. IV we mainly study two types of topological QPLs with a small external
magnetic field. The first one is $Z_{2}$ topological QPLs with the small
external magnetic field studied in Sec. IV A, and the other is the topological
chiral QPLs with the small external magnetic field investigated in Sec. IV B.
Finally, in Sec. V, we conclude our discussions.

\section{The generalized attractive Hubbard model at large-U limit}

Our starting point is the two dimensional generalized attractive Hubbard model
in a bipartite lattice (for example, the honeycomb lattice and the square
lattice), of which the Hamiltonian is given by
\begin{align}
\hat{H}  &  =-\sum \limits_{i\in A,j\in B}\left(  t_{ij}\hat{c}_{i,\uparrow
}^{\dagger}\hat{c}_{j,\uparrow}+t_{ij}^{\ast}\hat{c}_{i,\downarrow}^{\dagger
}\hat{c}_{j,\downarrow}+\mathrm{H.c.}\right) \nonumber \\
&  -\sum \limits_{i,j\in A/B}\left(  t_{ij}^{\prime}\hat{c}_{i,\uparrow
}^{\dagger}\hat{c}_{j,\uparrow}-t_{ij}^{\prime \ast}\hat{c}_{i,\downarrow
}^{\dagger}\hat{c}_{j,\downarrow}+\mathrm{H.c.}\right)  -U\sum \limits_{i}%
\hat{n}_{i\uparrow}\hat{n}_{i\downarrow}\nonumber \\
&  -\mu \sum \limits_{i,\sigma}\hat{c}_{i\sigma}^{\dagger}\hat{c}_{i\sigma
}-h\sum \limits_{i}\left(  \hat{c}_{i,\uparrow}^{\dagger}\hat{c}_{i,\uparrow
}-\hat{c}_{i,\downarrow}^{\dagger}\hat{c}_{i,\downarrow}\right)  .
\label{mod1}%
\end{align}
Here, $i=\left(  i_{x},i_{y}\right)  $ labels the lattice sites, \textrm{A}
and \textrm{B }denote \textrm{A}-sublattices and \textrm{B}-sublattices,
$\sigma=\uparrow$, $\downarrow$ are spin-indices, $t_{ij}$ are hopping
parameters between the sites on different sublattices, $t_{ij}^{\prime}$ are
hopping parameters between the site on same sublattices, $U$ is the strength
of the attractive interaction, $\mu$ is the chemical potential, and $h$ is the
strength of the Zeeman field. In the following parts, we consider the case
with $\mu=-U/2$, and set the lattice constant to be unity.

Let's discuss the global symmetry and the spontaneously symmetry breaking of
the original Hamiltonian in Eq.(\ref{mod1}). The Hamiltonian in Eq.(\ref{mod1}%
) has an \textrm{SU(2)} particle-hole (pseudo-spin) symmetry group when
$\mu=-U/2$, in which the \textrm{SU(2)} group elements act on the space of the
SF/CDW order parameters. To make the \textrm{SU(2)} pseudo-spin symmetry more
clear, we note that in terms of the canonical particle-hole
transformation\cite{yang}
\[
\hat{c}_{i,\uparrow}\rightarrow \tilde{c}_{i,\uparrow},\text{ }\hat
{c}_{i,\downarrow}\rightarrow(-1)^{i_{x}+i_{y}}\tilde{c}_{i,\downarrow}^{\dag
},
\]
the original model is mapped onto a repulsive Hubbard model with the effective
chemical potential $\tilde{\mu}=h+U/2$ and effective Zeeman field $\tilde
{h}=\mu+U/2$ as $\hat{H}\rightarrow \tilde{H}$ , where
\begin{align}
\tilde{H}  &  =-\sum \limits_{i\in A,j\in B}t_{ij}\tilde{c}_{i,\sigma}%
^{\dagger}\tilde{c}_{j,\sigma}-\sum \limits_{i,j\in A/B}t_{ij}^{\prime}%
\tilde{c}_{i,\sigma}^{\dagger}\tilde{c}_{j,\sigma}+\mathrm{H.c.}\nonumber \\
&  -\tilde{h}\sum \limits_{i,\alpha,\beta}\tilde{c}_{i,\alpha}^{\dagger}%
\sigma_{\alpha \beta}^{z}\tilde{c}_{i,\beta}-\tilde{\mu}\sum \limits_{i,\sigma
}\tilde{n}_{i,\sigma}+U\sum \limits_{i}\tilde{n}_{i,\uparrow}\tilde
{n}_{i,\downarrow}. \label{mod2}%
\end{align}
Note that we have omitted the constant term $-(\mu-h)N$ with $N$ the total
lattice sites number in Hamiltonian $\tilde{H}$. Then, for the case of
$\tilde{h}=\mu+U/2=0$, i.e., $\mu=-U/2$, we can define an \textrm{SU(2)}
pseudo-spin symmetry of the Hamiltonian in Eq.(\ref{mod1}), i.e.,
\begin{equation}
\hat{H}\rightarrow \hat{H}^{\prime}=\mathcal{U}\tilde{H}\mathcal{U}^{-1}%
=\tilde{H}%
\end{equation}
by doing a pseudo-spin rotation
\begin{equation}
\Psi \rightarrow \Psi^{^{\prime}}=\mathcal{U}\Psi,
\end{equation}
with $\Psi=(\tilde{c}_{i,\uparrow},\tilde{c}_{i,\downarrow})^{T}.$ The
\textrm{SU(2)} pseudo-spin operators of the attractive Hubbard model thus
become
\begin{gather}
\hat{\eta}^{-}\leftrightarrow \left(  -1\right)  ^{i_{x}+i_{y}}\hat{\Delta}%
_{i}=\left(  -1\right)  ^{i_{x}+i_{y}}\hat{c}_{i,\downarrow}\hat
{c}_{i,\uparrow},\text{ }\nonumber \\
\hat{\eta}^{+}\leftrightarrow \left(  -1\right)  ^{i_{x}+i_{y}}\hat{\Delta}%
_{i}^{\dag}=\left(  -1\right)  ^{i_{x}+i_{y}}\hat{c}_{i,\uparrow}^{\dag}%
\hat{c}_{i,\downarrow}^{\dag},\\
\hat{\eta}^{z}\leftrightarrow(\hat{\rho}_{i}-1)/2,\nonumber
\end{gather}
where $\hat{\eta}^{\pm}=\hat{\eta}^{x}\pm i\hat{\eta}^{y},$ $\hat{\rho}_{i}$
is the particle density operator, and the \textrm{SU(2)} algebraic relation
between the \textrm{SU(2)} pseudo-spin operators is $[\hat{\eta}^{\alpha},$
$\hat{\eta}^{\beta}]=i\epsilon_{\alpha \beta \gamma}\hat{\eta}^{\gamma}$. The
ground state with $\left \langle \hat{\eta}^{+}\right \rangle \neq0$ or
$\left \langle \hat{\eta}^{-}\right \rangle \neq0$ is an SF state and the ground
state with $\left \langle \hat{\eta}^{z}\right \rangle \neq0$ is a charge
density wave (CDW) state.

For the generalized attractive Hubbard model described by Eq.(\ref{mod1}),
with increasing the interaction strength, the ground state turns into a paired
state. At the large-$U$ limit, such paired state has a large energy gap about
$U$. For above generalized attractive Hubbard model with \textrm{SU(2)}
particle-hole (pseudo-spin) symmetry, by integrating out the hopping terms, we
derive an effective pseudo-spin model with the super-exchange terms
\begin{align}
H_{s}  &  =\sum \limits_{i\in A,j\in B}J_{ij}\mathbf{\hat{\eta}}_{i}%
\cdot \mathbf{\hat{\eta}}_{j}+\sum \limits_{i,j\in A/B}J_{ij}^{\prime
}\mathbf{\hat{\eta}}_{i}\cdot \mathbf{\hat{\eta}}_{j}\nonumber \\
&  =\sum \limits_{i\in A,j\in B}J_{ij}\left(  \hat{\eta}_{i}^{x}\hat{\eta}%
_{j}^{x}+\hat{\eta}_{i}^{y}\hat{\eta}_{j}^{y}+\hat{\eta}_{i}^{z}\hat{\eta}%
_{j}^{z}\right) \nonumber \\
&  +\sum \limits_{i,j\in A/B}J_{ij}^{\prime}\left(  \hat{\eta}_{i}^{x}\hat
{\eta}_{j}^{x}+\hat{\eta}_{i}^{y}\hat{\eta}_{j}^{y}+\hat{\eta}_{i}^{z}%
\hat{\eta}_{j}^{z}\right)  ,
\end{align}
where the super-exchange coupling constants between different and same
sublattices are $J_{ij}=4\left \vert t_{ij}\right \vert ^{2}/U$ and
$J_{ij}^{\prime}=4\left \vert t_{ij}^{\prime}\right \vert ^{2}/U$, respectively.
One can see that $H_{s}$ is really a frustrated pseudo-spin model.

For a conventional SF order with spontaneous \textrm{U(1)} phase symmetry
breaking, there exists one Goldstone mode with respect to the quantum phase
fluctuation. In two dimensions, there exists a Kosterlitz-Thouless (KT)
transition, below which the (quasi) long range phase coherence establishes.
Now we have an SF/CDW order with spontaneously \textrm{SU(2)} pseudo-spin
rotation symmetry breaking. Hence, the quantum fluctuations around the mean
field ground state are much stronger. In the CDW phase, one has a nonzero
particle density modulation at different sublattices $\left \langle \hat{\eta
}^{z}\right \rangle \neq0$. According to the commutation relation between the
phase $\phi_{i}$ of $\Delta_{i}$ and the particle density operator $\hat{\rho
}_{i}$, i.e., $[\phi_{i},\hat{\rho}_{i}]\neq0$, the nonzero particle density
modulation leads to an uncertainty for the SF phase coherence and even
destroys the long range SF phase coherence. Thus, the non-zero value of
$\Delta_{0}=\left \vert \Delta_{i}\right \vert $ only means the existence of
Cooper pairing. It does not necessarily imply that the ground state is a long
range SF order. As a result, one needs to examine the stability of the SF
order against quantum fluctuations based on a formulation by keeping
\textrm{SU(2)} pseudo-spin rotation symmetry.

\section{Projective space construction of QPLs}

For the frustrated pseudo-spin model $H_{s}$, the SF may possesses exotic
quantum orders and also cannot be described by Landau's symmetry breaking
theory. We call it QPL, of which the fermions are paired and single
quasi-particle's excitation has a finite energy gap. However, we don't have
long range SF phase coherence and the SF order parameter is zero, i.e.,
$\langle \hat{c}_{i\uparrow}^{\dag}\hat{c}_{i\downarrow}^{\dag}\rangle=0$. Thus
in this region, the SF correlation decays exponentially $\left \langle
\Delta^{\ast}(x,y)\Delta(0)\right \rangle \rightarrow0$. The particle number
must be $2$ or $0$ on each site that correspond to doublon or holon,
respectively. See the illustration in Fig.(1).
\begin{figure}[h]
\scalebox{0.65}{\includegraphics* [0.1in,1.0in][5.5in,3.5in]{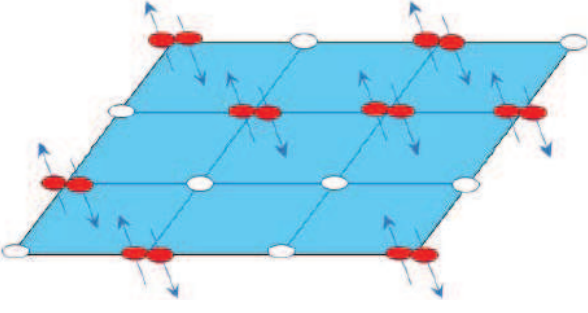}}\caption{The
illustration of the quantum phase liquid. The two-red-circle represents doublon, and the white-circle represents holon.}%
\label{Fig1}%
\end{figure}In this section, we are going to use the projective space
construction(PSG) to construct quantum phase liquids by the $SU(2)$
slave-boson approach\cite{wen,Wenpub}. The gauge structure here is just that
of the quantum spin liquid. The pseudo-spin operator $\hat{\eta}_{i}$ is now
represented as
\begin{equation}
\hat{\eta}_{i}=\frac{1}{2}\tilde{f}_{i\alpha}^{\dag}\sigma_{\alpha \beta}%
\tilde{f}_{i\beta} \label{sl2.2}%
\end{equation}
where $\tilde{f}$ denotes pseudo spinon. In terms of the fermion operators the
Hamiltonian $H_{s}$ can be rewritten as%
\begin{align}
H_{s}  &  =\sum \limits_{i\in A,j\in B}J_{ij}\mathbf{\hat{\eta}}_{i}%
\cdot \mathbf{\hat{\eta}}_{j}+\sum \limits_{i,j\in A/B}J_{ij}^{\prime
}\mathbf{\hat{\eta}}_{i}\cdot \mathbf{\hat{\eta}}_{j}\\
&  =-\frac{1}{2}\sum \limits_{i\in A,j\in B}J_{ij}\tilde{f}_{j\alpha}^{\dagger
}\tilde{f}_{i\alpha}\tilde{f}_{i\beta}^{\dagger}\tilde{f}_{j\beta}\nonumber \\
&  -\frac{1}{2}\sum \limits_{i,j\in A/B}J_{ij}^{\prime}\tilde{f}_{j\alpha
}^{\dagger}\tilde{f}_{i\alpha}\tilde{f}_{i\beta}^{\dagger}\tilde{f}_{j\beta
}\nonumber
\end{align}
with the constraints $\tilde{f}_{i\alpha}^{\dag}\tilde{f}_{i\alpha}=1$ and
$\tilde{f}_{i\alpha}\tilde{f}_{i\beta}\epsilon_{\alpha \beta}=0.$ Such
constraints can be enforced by including the site-dependent and the
time-independent Lagrangian multipliers: $a_{0}^{3}(i)(\tilde{f}_{i\alpha
}^{\dag}\tilde{f}_{i\alpha}-1)$, $(a_{0}^{1}+ia_{0}^{2})\tilde{f}_{i\alpha
}\tilde{f}_{i\beta}\epsilon_{\alpha \beta}$ in the Hamiltonian.

Replacing the operators $\tilde{f}_{i\alpha}\tilde{f}_{j\beta}$ and $\tilde
{f}_{i\alpha}^{\dag}\tilde{f}_{j\beta}$ by their mean field values
\begin{equation}
\Delta_{ij}=-2\left \langle \epsilon_{\alpha \beta}\tilde{f}_{i\alpha}\tilde
{f}_{j\beta}\right \rangle ,\text{ }\chi_{ij}=2\left \langle \delta_{\alpha
\beta}\tilde{f}_{i\alpha}^{\dag}\tilde{f}_{j\beta}\right \rangle ,
\label{sl2a.4}%
\end{equation}
we arrive at the mean-field Hamiltonian
\begin{align}
H_{\mathrm{mean}}  &  =-\frac{3}{8}\sum \limits_{i\in A,j\in B}J_{ij}(\chi
_{ji}\tilde{f}_{i\alpha}^{\dag}\tilde{f}_{j\alpha}+\Delta_{ij}\tilde
{f}_{i\alpha}^{\dag}\tilde{f}_{j\beta}^{\dag}\epsilon_{\alpha \beta
}+\mathrm{H.c.})\nonumber \\
&  -\frac{3}{8}\sum \limits_{i,j\in A/B}J_{ij}^{\prime}(\chi_{ji}\tilde
{f}_{i\alpha}^{\dag}\tilde{f}_{j\alpha}+\Delta_{ij}\tilde{f}_{i\alpha}^{\dag
}\tilde{f}_{j\beta}^{\dag}\epsilon_{\alpha \beta}+\mathrm{H.c.})\nonumber \\
&  +\frac{3}{8}\sum \limits_{i\in A,j\in B}J_{ij}\left(  |\chi_{ij}%
|^{2}+|\Delta_{ij}|^{2}\right) \nonumber \\
&  +\frac{3}{8}\sum \limits_{i,j\in A/B}J_{ij}\left(  |\chi_{ij}|^{2}%
+|\Delta_{ij}|^{2}\right) \nonumber \\
&  +\sum_{i}\{a_{0}^{3}(\tilde{f}_{i\alpha}^{\dag}\tilde{f}_{i\alpha
}-1)+[(a_{0}^{1}+ia_{0}^{2})\tilde{f}_{i\alpha}\tilde{f}_{i\beta}%
\epsilon_{\alpha \beta}+\mathrm{H.c.}]\}, \label{mean1}%
\end{align}
where $\chi_{ij}$, $\Delta_{ij}$ and $a_{0}^{l}$ can be derived by the
mean-field approach for a given QPL. Let $|\Psi_{\mathrm{mean}}\rangle$ be the
ground state of $H_{\mathrm{mean}}$. Then a many-body state can be obtained
from the mean-field state $|\Psi_{\mathrm{mean}}\rangle$ by projecting
$|\Psi \rangle=\mathcal{P}|\Psi_{\mathrm{mean}}\rangle$ into the subspace with
a single occupation.

In addition, we introduce $SU(2)$ doublet
\begin{equation}
\psi=\left(
\begin{array}
[c]{c}%
f_{1}\\
f_{2}%
\end{array}
\right)  =\left(
\begin{array}
[c]{c}%
\tilde{f}_{\uparrow}\\
\tilde{f}_{\downarrow}^{\dag}%
\end{array}
\right)
\end{equation}
and the matrix
\begin{equation}
U_{ij}=\left(
\begin{array}
[c]{cc}%
\chi_{ij}^{\dag} & \Delta_{ij}\\
\Delta_{ij}^{\dag} & -\chi_{ij}%
\end{array}
\right)  =U_{ji}^{\dag}. \label{sl2a.7}%
\end{equation}
Using them, we can rewrite the mean field Hamiltonian of the QPL and the
constraints into more compact formula as
\begin{align}
H_{\mathrm{mean}}  &  =\frac{3}{8}\sum \limits_{i\in A,j\in B}J_{ij}[\frac
{1}{2}\text{\textrm{Tr}}(U_{ij}^{\dag}U_{ij})-(\psi_{i}^{\dag}U_{ij}\psi
_{j}+\mathrm{H.c.})]\nonumber \\
&  +\frac{3}{8}\sum \limits_{i,j\in A/B}J_{ij}^{\prime}[\frac{1}{2}%
\text{\textrm{Tr}}(U_{ij}^{\dag}U_{ij})-(\psi_{i}^{\dag}U_{ij}\psi
_{j}+\mathrm{H.c.})]\nonumber \\
&  +\sum_{i}a_{0}^{l}\psi_{i}^{\dag}\tau^{l}\psi_{i} \label{mean}%
\end{align}
and
\begin{equation}
\left \langle \psi_{i}^{\dag}\tau^{l}\psi_{i}\right \rangle =0,
\end{equation}
where $\tau^{l}$ are the Pauli matrices with $l=1,2,3$.

Similar to the quantum spin liquids, the mean field Hamiltonian
$H_{\mathrm{mean}}$ of QPL is invariant under a local $SU(2)$ transformation
$W(i)$\cite{wen,Wenpub}:
\begin{equation}
\psi_{i}\rightarrow W(i)\psi_{i},\text{ }U_{ij}\rightarrow W(i)U_{ij}W^{\dag
}(j). \label{sl2a.9}%
\end{equation}
In particular, two mean-field ansatzs that have different $(U_{ij},a_{0}^{l})$
and $(U_{ij}^{\prime},a_{0}^{\prime l})$ may be the same QPL by an
$\mathrm{SU(2)}$ gauge transformation. To understand the gauge fluctuations
around the above mean-field state of a QPL, we may observe the non-trivial
$\mathrm{SU(2)}$ flux through plaquettes. These fluxes may break
$\mathrm{SU(2)}$ gauge structure down to $\mathrm{U(1)}$ or $Z_{2}$ gauge
structures and play the role of the Higgs fields. Thus we may characterize
different quantum phase liquids by different PSGs and different gauge
structures as people understand the quantum spin liquids.

\section{Topological QPLs with an external magnetic field}

Now we consider the generalized attractive Hubbard model with an external
magnetic field, of which the Hamiltonian becomes
\begin{align}
\hat{H}  &  =-\sum \limits_{i\in A,j\in B}e^{i(\theta_{i}-\theta_{j})}\left(
t_{ij}\hat{c}_{i,\uparrow}^{\dagger}\hat{c}_{j,\uparrow}+t_{ij}^{\ast}\hat
{c}_{i,\downarrow}^{\dagger}\hat{c}_{j,\downarrow}\right) \nonumber \\
&  -\sum \limits_{i,j\in A/B}e^{i(\theta_{i}-\theta_{j})}\left(  t_{ij}%
^{\prime}\hat{c}_{i,\uparrow}^{\dagger}\hat{c}_{j,\uparrow}-t_{ij}^{\prime
\ast}\hat{c}_{i,\downarrow}^{\dagger}\hat{c}_{j,\downarrow}\right)  -\mu
\sum \limits_{i,\sigma}\hat{n}_{i\sigma}\nonumber \\
&  -U\sum \limits_{i}\hat{n}_{i\uparrow}\hat{n}_{i\downarrow}-h\sum
\limits_{i}\left(  \hat{c}_{i,\uparrow}^{\dagger}\hat{c}_{i,\uparrow}-\hat
{c}_{i,\downarrow}^{\dagger}\hat{c}_{i,\downarrow}\right)  +\mathrm{H.c.}%
\end{align}
where the spatial variation of the phase $e^{i(\theta_{i}-\theta_{j})}$ can be
interpreted as an effective \textit{Aharonov-Bohm} phase induced by the
external magnetic field. With the choice of the symmetric gauge, we can define
the flux ratio $\alpha=\Phi/\Phi_{0}$ with $\Phi$ the magnetic flux through a
plaquette, and $\Phi_{0}=h/\left(  2e\right)  $ the half of flux quantum. The
sum of the phases along a closed loop surrounding the plaquette is $\theta
_{i}^{\prime}-\theta_{i}=\pi \alpha$, which is actually $\alpha$ $\pi$-flux per
plaquette. The strength of the external magnetic field is $B=\pi \alpha/S_{0}$,
where $S_{0}$ is the area of a plaquette. In this paper we only consider the
limit of a small external field, $B\rightarrow0$ ($\alpha \ll1$).

At the large-U limit, the generalized attractive Hubbard model with an
external magnetic field turns into an effective pseudo-spin model with
external magnetic field, of which the Hamiltonian is reduced into
\begin{align}
H_{s}  &  =J_{ij}\sum \limits_{i\in A,j\in B}(e^{i2(\theta_{i}-\theta_{j})}%
\eta_{i}^{+}\eta_{j}^{-}+e^{-i2(\theta_{i}-\theta_{j})}\eta_{i}^{-}\eta
_{j}^{+})\nonumber \\
&  +\sum \limits_{i\in A,j\in B}J_{ij}\eta_{i}^{z}\eta_{j}^{z}+J_{ij}^{\prime
}\sum \limits_{i,j\in A/B}(e^{i2(\theta_{i}-\theta_{j})}\eta_{i}^{+}\eta
_{j}^{-}\nonumber \\
&  +e^{-i2(\theta_{i}-\theta_{j})}\eta_{i}^{-}\eta_{j}^{+})+\sum
\limits_{i,j\in A/B}J_{ij}^{\prime}\eta_{i}^{z}\eta_{j}^{z}.
\end{align}
Note that in above equation we have used the definition of the \textrm{SU(2)}
pseudo-spin operators
\begin{gather}
\hat{\eta}^{-}=\left(  -1\right)  ^{i_{x}+i_{y}}\hat{c}_{i,\downarrow}\hat
{c}_{i,\uparrow},\text{ }\hat{\eta}^{+}=\left(  -1\right)  ^{i_{x}+i_{y}}%
\hat{c}_{i,\uparrow}^{\dag}\hat{c}_{i,\downarrow}^{\dag},\\
\hat{\eta}^{z}=(\hat{\rho}_{i}-1)/2.\nonumber
\end{gather}
Then, we find that the mean-field Hamiltonian of slave-particles becomes
\begin{align}
H_{\mathrm{mean}}  &  \rightarrow H_{\mathrm{mean}}^{\prime}=-\frac{3}{8}%
\sum \limits_{i\in A,j\in B}J_{ij}(\chi_{ji}e^{2i(\theta_{i}-\theta_{j})}%
\tilde{f}_{i\alpha}^{\dag}\tilde{f}_{j\alpha}\nonumber \\
&  +\Delta_{ij}e^{2i(\theta_{i}+\theta_{j})}\tilde{f}_{i\alpha}^{\dag}%
\tilde{f}_{j\beta}^{\dag}\epsilon_{\alpha \beta})\nonumber \\
&  -\frac{3}{8}\sum \limits_{i,j\in A/B}J_{ij}^{\prime}(\chi_{ji}%
e^{2i(\theta_{i}-\theta_{j})}\tilde{f}_{i\alpha}^{\dag}\tilde{f}_{j\alpha
}\nonumber \\
&  +\Delta_{ij}e^{2i(\theta_{i}+\theta_{j})}\tilde{f}_{i\alpha}^{\dag}%
\tilde{f}_{j\beta}^{\dag}\epsilon_{\alpha \beta})+\frac{3}{8}\sum \limits_{i\in
A,j\in B}\left(  |\chi_{ij}|^{2}+|\Delta_{ij}|^{2}\right) \nonumber \\
&  +\frac{3}{8}\sum \limits_{i,j\in A/B}\left(  |\chi_{ij}|^{2}+|\Delta
_{ij}|^{2}\right)  +\sum_{i}\{a_{0}^{3}(\tilde{f}_{i\alpha}^{\dag}\tilde
{f}_{i\alpha}-1)\nonumber \\
&  +[(a_{0}^{1}+ia_{0}^{2})\tilde{f}_{i\alpha}\tilde{f}_{i\beta}%
\epsilon_{\alpha \beta}]\}+\mathrm{H.c.}.
\end{align}

\begin{figure}[h]
\scalebox{0.52}{\includegraphics* [0.1in,0.3in][11.7in,3.0in]{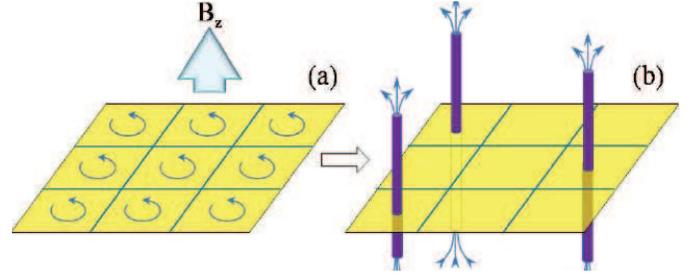}}\caption{The
illustration of the induced $\pi$-flux by external field in the topological
quantum phase liquid: the uniform magnetic flux in each plaquette turns into a
gas of $\pi$-flux.}%
\label{Fig2}%
\end{figure}

From this effective Hamiltonian of the slave-particles, it can be seen that
the slave-particles $\tilde{f}_{i\alpha}$ see $2\alpha$ $\pi$-flux per
plaquette. While, for the quantum spin liquid, the situation is much
different: the external magnetic field never changes the effective spin model
and the mean-field Hamiltonian of slave-particles. Since quantum phase liquids
with topological order are protected by the finite energy gaps of excitations,
they are stable against arbitrary local perturbations\cite{wen,Wenrig,Wtoprev}%
. As a result, for the topological phase liquids, we can assume that the small
external magnetic field will not change their mean field ansatz and the
corresponding PSG of the topological order. Hence, for the topological phase
liquids, since the mean-field Hamiltonian of slave-particles in Eq.(\ref{mean}%
) changes the original PSG, it is not quite right. For this case, the external
magnetic field will induce quantized vortices. In the topological phase
liquids in a lattice, the induced vortex must be $\pi$-flux and the density of
the induced quantized vortices $\rho_{v}$ is $BS_{0}/\pi$. See Fig.(2). The
true effective Hamiltonian of the topological phase liquids with a finite
external magnetic field is given by
\begin{align}
H_{\mathrm{mean}}  &  \rightarrow H_{\mathrm{mean}}^{\prime}\nonumber \\
&  =-\frac{3}{8}\sum \limits_{i\in A,j\in B}J_{ij}(\chi_{ji}\tilde{f}_{i\alpha
}^{\dag}\tilde{f}_{j\alpha}+\Delta_{ij}\tilde{f}_{i\alpha}^{\dag}\tilde
{f}_{j\beta}^{\dag}\epsilon_{\alpha \beta}+\mathrm{H.c.})\nonumber \\
&  +\frac{3}{8}\sum \limits_{i\in A,j\in B}J_{ij}\left(  |\chi_{ij}%
|^{2}+|\Delta_{ij}|^{2}\right) \nonumber \\
&  -\frac{3}{8}\sum \limits_{i,j\in A/B}J_{ij}^{\prime}(\chi_{ji}\tilde
{f}_{i\alpha}^{\dag}\tilde{f}_{j\alpha}+\Delta_{ij}\tilde{f}_{i\alpha}^{\dag
}\tilde{f}_{j\beta}^{\dag}\epsilon_{\alpha \beta}+\mathrm{H.c.})\nonumber \\
&  +\frac{3}{8}\sum \limits_{i,j\in A/B}J_{ij}\left(  |\chi_{ij}|^{2}%
+|\Delta_{ij}|^{2}\right) \nonumber \\
&  +\sum_{i}\{a_{0}^{3}(\tilde{f}_{i\alpha}^{\dag}\tilde{f}_{i\alpha
}-1)+[(a_{0}^{1}+ia_{0}^{2})\tilde{f}_{i\alpha}\tilde{f}_{i\beta}%
\epsilon_{\alpha \beta}]+\mathrm{H.c.}\} \nonumber \\
&  +H_{\mathrm{vortex}},
\end{align}
where $H_{\mathrm{vortex}}$ is the Hamiltonian of the induced quantized
vortices ($\pi$-flux). In the following parts we will discuss two kinds of
topological phase liquids with a finite density of quantized vortices.

\subsection{$Z_{2}$ topological QPLs with a small external magnetic field}

In this section we will discuss $Z_{2}$ topological QPLs with a small external
magnetic field. $Z_{2}$ topological QPLs have the simplest topological order -
$Z_{2}$ topological order\cite{RS9173,Wsrvb}, of which the ground-state has
topological degeneracy\cite{wen4,Wsrvb}. The low energy effective theory for
those $Z_{2}$ topologically ordered states is a $Z_{2}$ gauge theory with
mutual $\pi$ statistics between quasi-particles, the fermions and the bosonic
$Z_{2}$-vortices (quantized vortices with $\pi$-flux on a plaquette). That
means the fermions will obtain $\pi$ phase after moving around $Z_{2}$-vortex.
For the $Z_{2}$ topological QPLs with a small external magnetic field, there
exist the induced $Z_{2}$-vortices with the density $B/\pi$ (Here and in the
following, we assume a plaquette area $S_{0}\equiv1$).

It has been pointed out that the continuum effective theory of two dimensional
$Z_{2}$ gauge theory is the \textrm{U(1)}$\times$\textrm{U(1)} mutual
Chern-Simons (MCS) theory\cite{kou1,kou}. In the \textrm{U(1)}$\times$
\textrm{U(1)} MCS theory, there are two types of auxiliary gauge fields
$A_{\mu}\ $and $a_{\mu}$ coupled to $Z_{2}$ vortices and fermions
respectively. A mutual Chern-Simons term is introduced between $A_{\mu}\ $and
$a_{\mu}$ as follows,
\begin{equation}
\mathcal{L}_{MCS}=\frac{i}{\pi}\epsilon^{\mu \nu \lambda}A_{\mu}\partial_{\nu
}a_{\lambda}.
\end{equation}
Including the induced $Z_{2}$-vortex by the external magnetic fluxes, the
effective continuum theory of the $Z_{2}$ topological QPLs becomes
\begin{equation}
\mathcal{L}_{\mathrm{eff}}=\mathcal{L}_{f}+\mathcal{L}_{v}+\mathcal{L}_{MCS},
\label{mCSL}%
\end{equation}
where $\mathcal{L}_{f}$ is the effective lagrangian of the fermions and
$\mathcal{L}_{v}$ is the Lagrangian of $Z_{2}$-vortex
\begin{equation}
\mathcal{L}_{v}=\varphi^{\ast}\left[  \partial_{0}-ia_{0}-\mu_{v}\right]
\varphi+\varphi^{\ast}\frac{\left(  -i\partial_{\alpha}-a_{\alpha}\right)
^{2}}{2m_{v}}\varphi. \label{continous}%
\end{equation}
Here, $\varphi$ is the field of the bosonic $Z_{2}$-vortices, and $m_{v}$ is
the mass of $Z_{2}$-vortices.

For the dilute gas of the bosonic $Z_{2}$-vortices, the ground state becomes a
Bose-Einstein condensation (BEC) state with
\begin{equation}
\langle0|\varphi_{\mathbf{k}=0}|0\rangle=\sqrt{\rho_{v}}e^{i\phi}\neq0.
\end{equation}
Here, $|0\rangle$ denotes the ground state. Thus the ground state$\ $%
spontaneously breaks both \textrm{U(1)}$\times$ \textrm{U(1)} gauge symmetry.
With $\left \langle \varphi \right \rangle \neq0$, $\mathcal{L}_{v}$ in
(\ref{continous}) reduces to
\begin{equation}
\mathcal{L}_{v}=i\rho_{v}(\partial_{0}\phi-a_{0}+i\mu_{v})+\frac{\rho_{v}%
}{2m_{v}}\left(  \mathbf{\nabla}\phi-\mathbf{a}\right)  ^{2}%
\end{equation}
\newline by writing $\varphi(\mathbf{r})=\sqrt{\rho_{v}}e^{i\phi(\mathbf{r})}%
$. Then the mutual Chern-Simons term (\ref{continous}) can be rewritten as
\begin{equation}
\mathcal{L}_{MCS}=-\frac{i}{\pi}\mathbf{a}\cdot(\mathbf{E}\times
\hat{\mathbf{z}})+\frac{i}{\pi}a_{0}\mathbf{B}\cdot \hat{\mathbf{z}}
\label{cs-cs}%
\end{equation}
by introducing the \textquotedblleft electric\textquotedblright \ field
$\mathbf{E}=\partial_{0}\mathbf{A}-\nabla A_{0}$ and \textquotedblleft
magnetic\textquotedblright \ field $\mathbf{B}=$ $\nabla \times \mathbf{A}$ for
the vector potential $\mathbf{A}$.

First, integrating out the gauge field $a_{0}$, one obtains the condition%
\[
B\equiv B^{z}=\mathbf{B}\cdot \hat{\mathbf{z}}=\pi \rho_{v}%
\]
which is uniform and fixes the spatial component $\mathbf{A}$, such that
$\mathbf{E=}-\nabla A_{0}$. After integrating out $\mathbf{a},$ the resulting
effective Lagrangian takes the following form
\begin{align}
\mathcal{L}_{\mathrm{eff}}  &  =\mathcal{L}_{\mathrm{f}}+(\frac{m_{v}}{2\pi
B})\left \vert \mathbf{E}\right \vert ^{2}-\frac{i}{\pi}\epsilon^{0\nu \lambda
}A_{0}\partial_{\nu}\partial_{\lambda}\phi \nonumber \\
&  +\frac{\rho_{v}}{2m_{v}}\left(  \mathbf{\nabla}\phi \right)  ^{2}-\mu
_{v}\rho_{v}. \label{eff-sc}%
\end{align}
Next, we integrate out $A_{0}$ in Eq.(\ref{eff-sc}) and obtains the following
effective action in (2+1)-dimensional Euclidean space\cite{kou1}%
\[
S_{\mathrm{eff}}=\int d^{3}x_{\mu}\left[  \mathcal{L}_{f}\left(
a_{0}=0\right)
\right]  +\int dx_{0}V_{\mathrm{SC}},
\]
where
\begin{align}
V_{\mathrm{SC}}  &  =q_{s}^{2}\int d^{2}\mathbf{r}d^{2}\mathbf{r}^{\prime}%
\ln \left \vert \mathbf{r}-\mathbf{r}^{\prime}\right \vert \left[  \rho
_{f}(\mathbf{r})+\frac{1}{\pi}\epsilon^{0\nu \lambda}\partial_{\nu}%
\partial_{\lambda}\phi(\mathbf{r})\right] \nonumber \\
&  \times \left[  \rho_{f}(\mathbf{r}^{\prime})+\frac{1}{\pi}\epsilon
^{0\nu \lambda}\partial_{\nu}\partial_{\lambda}\phi(\mathbf{r}^{\prime
})\right]
\end{align}
with $q_{s}^{2}=\frac{\pi \rho_{v}}{4m_{v}}$. $\rho_{f}(\mathbf{r})$ denotes
the density of the fermionic spinons. So, there exists a Kosterlitz-Thouless
(KT) transition temperature\cite{BKT,kost1}%
\begin{equation}
k_{B}T_{\mathrm{KT}}\simeq \frac{q_{s}^{2}}{4}=\frac{\pi \rho_{v}}{16m_{v}}.
\end{equation}
When the temperature is lower than $T_{\mathrm{KT}},$ the fermionic spinons
are confined; when the temperature is higher than $T_{\mathrm{KT}}$, the
fermionic spinons can be free.

In particular, the BEC state of the $Z_{2}$-vortices is a super-solid state.
Let us calculate the "charge" conductance $\sigma_{c}$ of the system. The
origin of the dissipation in this BEC state is the flow of the $Z_{2}%
$-vortices as respondence to an external "electric" field. Using the relation
between vortex conductance $\sigma_{v}$ and and "charge" conductance
$\sigma_{c}$\cite{stone,wen5,kou1}:%
\begin{equation}
\sigma_{v}\sigma_{c}=\frac{1}{\pi^{2}},
\end{equation}
we can find that the zero vortex conductance $\sigma_{v}=\infty$ leads to a
diverge "charge" conductance $\sigma_{c}\rightarrow0$. That means $Z_{2}$
topological QPLs with a small external magnetic field is really an insulating
super-solid state. In addition, there exists anomalous Nernst effect. The
Nernst effect refers to a transverse "electric" field induced by applying a
temperature gradient on the system\cite{ner}. The mechanism leading to a
significant Nernst signal is the $Z_{2}$-vortex-flow in the topological QPL
phase. In the BEC state of the $Z_{2}$-vortices, the Nernst signal will also diverge.

\subsection{Topological chiral QPLs with a small external magnetic field}

Another topological QPL is chiral phase liquid (CPL) breaking time reversal
symmetry, of which the elementary excitations are anyons with fractional
statistics \cite{wen3}. To calculate the topological invariants in momentum
space of the topological chiral phase liquid, the above formulation in
Eq.(\ref{mod2}) is reduced into,
\begin{equation}
H_{\mathrm{mean}}=\sum_{k}\psi_{k}^{\dag}[\mathbf{u}(k)\cdot \mathbf{\tau]}%
\psi_{k}+h.c.
\end{equation}
where $\psi_{k}=(%
\begin{array}
[c]{cc}%
\tilde{f}_{1,k} & \tilde{f}_{2,k}%
\end{array}
)^{T}\ $and $\hat{\mathbf{u}}(k)=\mathbf{u}/|\mathbf{u}|$ is the unit vector
in the momentum space. By introducing the Chern number
\begin{equation}
\mathcal{C}={\frac{1}{4\pi}}\int dk_{x}dk_{y}~\hat{\mathbf{u}}\cdot
({\frac{\partial \hat{\mathbf{u}}}{\partial{k_{x}}}}\times{\frac{\partial
\hat{\mathbf{u}}}{\partial{k_{y}}}),}%
\end{equation}
we may has a topological CPL with the non-zero Chern number $\mathcal{C}$. The
low energy effective theory for the the topological chiral phase liquid is the
Chern-Simons (CS) theory\cite{redlich,cs}
\begin{equation}
\mathcal{L}_{\mathrm{CS}}=\frac{i\mathcal{C}N}{4\pi}\epsilon^{\mu \nu \lambda
}a_{\mu}\partial_{\nu}a_{\lambda} \label{ChernSimons}%
\end{equation}
where $N$ is the flavor number of the fermionic spinons. Here $a_{\mu}$ is the
the auxiliary $\mathrm{U(1)}$ gauge fields. Another important property of CPL
is the non-zero chiral order parameter which is defined by\cite{wen3}
\begin{equation}
\mathcal{\chi}_{_{\langle123\rangle}}=\left \langle {\mathbf{\eta}}_{1}%
\cdot({\mathbf{\eta}}_{2}\times{\mathbf{\eta}}_{3})\right \rangle \neq0.
\end{equation}

For the topological CPL with a small external magnetic field, there also exist
the induced vortices with the density $B/\pi$. In the CPL, each quantized
vortex carries the "charge" number $\mathcal{C}N/2$ due to the Chern-Simon
term $\mathcal{L}_{\mathrm{CS}}$. Consequently, a quantized vortex becomes an
anyon. To calculate the many-body system of anyons, we will use a dual
description where we introduce a $\mathrm{U(1)}$ gauge field $b_{\mu}$ to
describe the density $j^{0}$ and current $j^{i}$ of the bosonic field
$\varphi$ denoting anyons: $j^{\mu}=\frac{i}{2\pi}\epsilon^{\mu \nu \lambda
}\partial_{\nu}b_{\lambda}$. Now the effective continuum Lagrangian of the
topological CPL with a small external magnetic field turns into%
\begin{equation}
\mathcal{L}_{\mathrm{eff}}=\mathcal{L}_{v}+\mathcal{L}_{f}+\mathcal{L}%
_{\mathrm{CS}}+\frac{i}{2\pi}\epsilon^{\mu \nu \lambda}b_{\mu}\partial_{\nu
}a_{\lambda},
\end{equation}
where $\mathcal{L}_{f}$ is the effective lagrangian of the fermions coupling
to $a_{\lambda}$, and $\mathcal{L}_{v}$ is the Lagrangian of anyons
\begin{equation}
\mathcal{L}_{v}=\varphi^{\ast}D_{0}\varphi+\frac{\left \vert D\varphi
\right \vert ^{2}}{2m_{v}}-\mu_{v}\varphi^{\ast}\varphi.
\end{equation}
Here, the bosonic field $\varphi$ denotes anyons that couple to both $a_{\nu}$
and $b_{\nu}$, $D_{\nu}=\partial_{\nu}-ia_{\nu}/2-ib_{\nu}$ is the covariant
derivative. $m_{v}$ is the mass of anyons.

The ground state of the dilute anyon gas is really an anyon superfluids. To
obtain the universal features from the anyon superfluid, we define
\[
a_{+,\mu}=b_{\mu}+\frac{a_{\mu}}{2},\ a_{-,\mu}=b_{\mu}-\frac{a_{\mu}}{2}%
\]
(or $a_{\mu}=a_{+,\mu}-a_{-,\mu}$, $b_{\mu}=(a_{+,\mu}+a_{-,\mu})/2$), and
rewrite the effective lagrangian as
\begin{align}
\mathcal{L}_{\mathrm{eff}} &  =\mathcal{L}_{f}+\varphi^{\ast}D_{0}%
\varphi+\frac{\left \vert D\varphi \right \vert ^{2}}{2m_{v}}+\mu_{v}%
\varphi^{\ast}\varphi \nonumber \\
&  +\frac{i(\mathcal{C}N+1)}{4\pi}a_{+,\mu}\partial_{\nu}a_{+,\lambda}%
\epsilon^{\mu \nu \lambda}\nonumber \\
&  +\frac{i(\mathcal{C}N-1)}{4\pi}a_{-,\mu}\partial_{\nu}a_{-,\lambda}%
\epsilon^{\mu \nu \lambda}-\frac{i\mathcal{C}N}{2\pi}a_{+,\mu}\partial_{\nu
}a_{-,\lambda}\epsilon^{\mu \nu \lambda}.
\end{align}
At zero temperature, with $\left \langle \varphi \right \rangle =\sqrt{\rho_{v}%
}e^{i\phi(\mathbf{r})}\neq0,$ we get%
\begin{equation}
\mathcal{L}_{v}=i\rho_{v}(\partial_{0}\phi-a_{+,0}-\mu_{v})+\frac{\rho_{v}%
}{2m_{v}}\left(  \mathbf{\nabla}\phi-\mathbf{a}_{+}\right)  ^{2}.
\end{equation}
After integrating $a_{+,\mu}$\cite{note1}, we get%
\begin{align}
\mathcal{L}_{\mathrm{eff}} &  =\mathcal{L}_{\mathrm{f}}+\frac{i\left(
\mathcal{C}N-1\right)  }{4\pi}a_{-,\mu}\partial_{\nu}a_{-,\lambda}%
\epsilon^{\mu \nu \lambda}+\frac{\rho_{v}}{2m_{v}}\left(  \mathbf{\nabla}%
\phi \right)  ^{2}\nonumber \\
&  +\frac{m_{v}\mathcal{C}N}{4\pi B^{c}}|\mathbf{E}_{-}\mathbf{|}^{2}%
-\frac{i\mathcal{C}N}{2\pi}\epsilon^{0\nu \lambda}a_{0,-}\partial_{v}%
\partial_{\lambda}\phi-\mu_{v}\rho_{v}%
\end{align}
where $\mathbf{E}_{-}=\partial_{0}\mathbf{a}_{-}-\mathbf{\nabla}a_{-,0}$ and
$B^{c}=2\pi \rho_{v}/\mathcal{C}N$.

When $\mathcal{C}N=1$, the situation is similar to that of $Z_{2}$ topological
phase liquids. There exists a Kosterlitz-Thouless (KT) transition
temperature\cite{BKT,kost1}%
\begin{equation}
k_{B}T_{\mathrm{KT}}\simeq \frac{q_{s}^{2}}{4}=\frac{\pi \rho_{v}}{16m_{v}}.
\end{equation}
When the temperature is lower than $T_{\mathrm{KT}}$, the fermionic spinons
are confined, and when the temperature is higher than $T_{\mathrm{KT}}$, the
fermionic spinons can be free. When $\mathcal{C}N\neq1$, the gauge field has
an energy gap, which implies that there is no phase transition at finite
temperature and the fermionic spinons are deconfined at all temperature.

\section{Conclusion}

We have studied the two dimensional generalized attractive Hubbard model in a
bipartite lattice, and by particle-hole transformation, we find the model has
$\mathrm{SU(2)}$ pseudo spin rotation symmetry when the chemical potential
$\mu=U/2$. With a further analysis, we observe that the nonzero particle
density modulation leads to an uncertainty for the SF phase coherence, and
even destroys the long range SF phase coherence, which corresponds to the
"quantum phase liquid". In analogy to quantum spin liquid, we make
a\ projective space construction of quantum phase liquids by the
$\mathrm{SU(2)}$ slave-boson approach. Then, we have shown two types of
topological phase liquids with a small external magnetic field - $Z_{2}$ QPLs
and CPLs.

In the end, we show the comparison between the quantum spin liquid and quantum
phase liquid in Table. (1).

\begin{widetext}
\begin{table*}[t]
\caption{The comparison between quantum spin liquid (QSL) and quantum phase liquid (QPL)}
\begin{tabular}
{ccccccc} \hline \hline & Definition & Symmetry & Classification & Excitation & External fields\\ \hline
QSL& Short range spin order & SU(2) spin rotation symmetry & PSG & Spinon&---&\\
QPL& Short range superfluid order &SU(2) particle-hole symmetry    & PSG & Doublon/Holon&Finite density of $\pi$ flux\\ \hline \hline
\end{tabular}
\end{table*}
\end{widetext}

\begin{acknowledgments}
This work is supported by National Basic Research Program of China (973
Program) under the grant No. 2011CB921803, 2012CB921704 and NFSC Grant No. 11174035.
\end{acknowledgments}


\begin{thebibliography}{99}
\bibitem {note1}In the process of integration of gauge field $\mathbf{a}_{+}$,
we only consider the term quadratic term $\frac{\rho_{v}}{2m_{v}}%
\mathbf{a}_{+}^{2}$ for $\mathbf{a}_{+}$, and negelct the term $\frac
{-i\left(  \mathcal{C}N+1\right)  }{4\pi}\mathbf{a}_{+}\mathbf{\cdot}\left(
\mathbf{E}_{+}\mathbf{\times z}\right)  $ due to that this term corresponding
to a low energy part.
\end{thebibliography}
\end{document}